\documentclass[11pt]{article}
\usepackage{graphicx, color, transparent}
\usepackage{amsfonts,amsmath, amssymb, undertilde, appendix, hyperref}
\author{Anton van Niekerk\\
\textit{Perimeter Institute for Theoretical Physics,}\\
\textit{\ Waterloo, Ontario N2L 2Y5, Canada}\\
\textit{Department of Physics \& Astronomy and Guelph-Waterloo Physics Institute,}\\
\textit{\ University of Waterloo, Waterloo, Ontario N2L 3G1, Canada}}
\title{Entanglement Entropy in NonConformal Holographic  Theories}

\begin{document}
\maketitle
\begin{abstract}
The entanglement entropy of various geometries is calculated for the boundary theory dual to a
 stack of $N$ $Dp$-branes.  The entanglement entropies are readily expressed in terms of the effective coupling at the appropriate
 energy scales.  The results are also compared to the entropy density of a black brane and some universal properties of holographic 
entropy are found.
\end{abstract}
\newpage
\tableofcontents
\newpage

\section{Introduction}

In Maldacena's AdS/CFT conjecture of 1997 the bulk spacetime of type IIB superstring theory with $N$ $D3$-branes has a geometry 
of AdS$_5 \times S^5$ in the large $N$ limit of the boundary CFT. See \cite{maldacena1} - \cite{aharony} and references for a 
comprehensive review of the
 subject and \cite{mcgreevy} - \cite{malchapter} for more recent reviews.  The generalization of this duality exists
 \cite{maldacena} when the boundary space is a stack of $Dp$-branes for $p \neq 3$. When $N \rightarrow \infty$ 
and $Ng^2_{YM}$ is large the gauge theory living on the boundary space is dual to a classical supergravity solution,
 however the gravity solution will no longer be anti-de Sitter space.  In this limit the bulk spacetime acquires 
a classical dilaton field and has a nonconstant curvature.  Both of these quantities must be small for the duality to hold.
  This will constrain the depth inside the bulk for which the duality is valid and in turn constrain
 the sizes of regions for which we can holographically compute the entanglement entropy, as we will discuss. 
  
Given a quantum system with a density matrix $\rho$ on complementary spatial regions $A$ and $B$, 
we can trace out its degrees of freedom in $B$ and find the reduced density matrix $\rho_A$ describing the system on $A$, by
\begin{eqnarray}
 \rho_A &=& \textrm{tr}_B \rho \nonumber \\
&\equiv& \sum_{\lvert\lambda\rangle\in B}\langle\lambda\rvert \rho \lvert\lambda\rangle.
\end{eqnarray}
The loss of information about the system in region $B$ 
allows us to define the von Neumann entropy 
\begin{equation}
 S_A = -\textrm{tr}_A \rho_A \log \rho_A,
\end{equation}
known as the entanglement entropy (EE) of $A$.

\begin{figure}
 \def\svgwidth{10cm}
\centering
\input{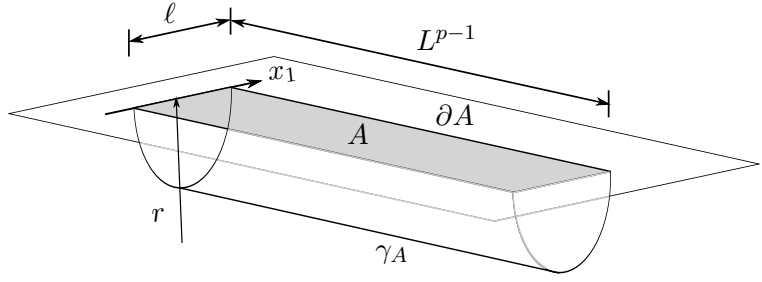}
\caption{The infinite strip A on the $Dp$-brane boundary of the $10$-dimensional bulk space, with the minimal
surface $\gamma_A$ in the bulk ending on $\partial A$. Also shown is the length regulator $L$ of the
$p-1$ lengthwise directions of A, as well as the $1$-dimensional width $\ell$ of the strip in the $x_1$-direction. The radial bulk 
direction is indicated by $r$.}\label{fig:dpbranestrip}
\end{figure}

It is a well-known fact that a black hole has entropy and that this entropy is proportional to the surface area of its event 
horizon \cite{bh}, \cite{hawk}.
  The entropy is given by the equation
\begin{equation}
 S_{BH} = \frac{A_{BH}}{4G_N},
\end{equation}
$A_{BH}$ being the area of the black hole's event horizon and $G_N$ being Newton's constant.
Ryu and Takayanagi made an important extension to this idea in \cite{rt}.  The extension was that the entanglement entropy
of a region $A$ in a CFT in $(p + 1)$-dimensions could be calculated by the
area of a minimal surface $\gamma_A$ in the $(p+2)$-dimensional AdS spacetime, using the AdS/CFT correspondence.  
$A$ is a $p$-dimensional region on the boundary of AdS space, with a 
$(p-1)$-dimensional boundary $\partial A$ on a fixed time-slice (see figures \ref{fig:dpbranestrip} and \ref{fig:dpbranedisk}), while
  $\gamma_A$ is the $p$-dimensional minimal
 surface connecting to $\partial A$.  Then the holographic EE of $A$ in the boundary CFT is
\begin{equation}
 S_A =  \min_{\partial\gamma_A=\partial A}\frac{A_{\gamma_A}}{4G^{(p+2)}_N}. \label{eq:area}
\end{equation}

When $A$ is a region on a $Dp$-brane, its EE can be calculated by \cite{stripdisk}
\begin{equation}
 S_A = 
\min_{\partial\gamma_A=\partial A}\frac{1}{4 G^{(10)}_N}\int_{\gamma_A}d^8 x\sqrt{-g}e^{-2\phi}. \label{eq:pent}
\end{equation}
$g$ is the determinant of the induced string frame metric on $\gamma_A$ and $G^{(10)}_N$ is Newton's constant in the bulk defined by
\begin{equation}
  16\pi G^{(10)}_N = (2\pi)^7 \alpha^{\prime 4}g^2_s. \label{eq:newton}
\end{equation}
 Notice that the integral (\ref{eq:pent}) for the entropy is $8$-dimensional.  This is because there 
are $9-p$ extra dimensions in the bulk compared to the $p$-dimensional boundary space.  The interpretation of 
$\gamma_A$ being a $p$-dimensional surface in the bulk is consistent, because $\gamma_A$ has rotational symmetry
in the $8-p$ additional spherical coordinates in the bulk and therefore the
 surface ``wraps'' these spherical directions.

The entanglement entropy of a region $A$ will diverge in the UV.  After regularization the EE can be expressed in terms of
a power series in the divergent parameter.  The leading order divergence will always have a coefficient proportional
 to the area of the boundary of $A$. In general, the coefficients of the polynomial orders of the UV cut-off will not occur in other 
regularization schemes, while the coefficient of a logarithmic term of the cut-off will be universal, meaning that it will
 occur in the EE independent of the regularization used.

Previous papers have calculated the EE of regions on $Dp$-branes.  See \cite{stripdisk} for a $D2$-brane,
 \cite{klebanov}, \cite{pakman}, \cite{d3}, \cite{d4} for $D3$ or $D4$-branes on orbifolds and \cite{Asplund:2011cq} for
 the $D1$-$D5$-brane
 system.
  In this paper the $Dp$-branes will not be orbifolded and 
the entanglement entropy will be expressed in a novel way in terms of the effective coupling, 
possibly giving a more intuitive form of the answer.

The rest of the paper is organized as follows:  In section 2 we introduce the metric 
and other details of the $10$-dimensional supergravity solution, as  well as the effective coupling and the high
energy regulators to be used in the sections that follow.
In section $3$ we evaluate the entanglement entropy of a strip and disk on the $Dp$-brane boundary of the $10$-dimensional
 bulk space in terms of the effective
 couplings and appropriate energy scales of the gauge theory. 
In section $4$ the results of section $3$ are discussed.  In the appendix
the entropy of a black hole in the supergravity solution is calculated, so that it can be compared with the 
entanglement entropies of the strip and disk.

\section{The supergravity solution}

The string frame supergravity solution describing the $Dp$-brane throat has a metric \cite{maldacena}, \cite{conventions}
\begin{equation}
 ds^2 = \left(\frac{r}{r_p}\right)^{\frac{7-p}{2}}\left(-dt^2 + \sum^{p}_{n=1}dx^2_n\right) 
+ \left(\frac{r_p}{r}\right)^{\frac{7-p}{2}}(dr^2
+ r^2 d\Omega^2_{8-p}) \label{eq:supgmetric}
\end{equation}
and a dilaton field
\begin{equation}
 e^{\phi} = \left(\frac{r_p}{r}\right)^{\frac{(7-p)(3-p)}{4}}, \label{eq:dil}
\end{equation}
where
\begin{eqnarray}
 r^{7-p}_p &=& Ng^{2}_{YM} \alpha^{\prime 5-p} d_p \label{eq:zp}\\
d_p &=& 2^{7-2p}\pi^{\frac{9-3p}{2}}\Gamma\left(\frac{7-p}{2}\right). \label{eq:dp}
\end{eqnarray}
$N$ is the rank of the $U(N)$ gauge group and $g^{2}_{YM}$ is the Yang-Mills coupling constant of the gauge theory
 defined by
\begin{equation}
  g^{2}_{YM} = (2\pi)^{p-2}g_s \alpha^{\prime \frac{p-3}{2}} \label{eq:gym}.
\end{equation}
In this coordinate system the boundary of the spacetime is located at $r = \infty$. The gauge theory is restricted to the 
$t$ and $x_n$ spacetime directions, while the radial direction $r$ is roughly related to the energy scales in the gauge theory.
Notice in equation (\ref{eq:gym}) that $g^{2}_{YM}$ will therefore have units of length to the power $p-3$.  Of course for $p=3$, 
(\ref{eq:supgmetric}) is the metric for AdS$_5 \times S^5$ and $g^{2}_{YM}$ will be dimensionless.
  We therefore obtain the AdS/CFT correspondence in this specific case.

Two further conventions that we will use are the energy regulator and the effective coupling at a certain energy scale.  
In calculating the holographic entanglement entropy, 
the area of $\gamma_A$ will be divergent close to the boundary of the bulk space (since the boundary lies at infinity).

The UV regulator $\Lambda$ in the gauge theory can be expressed as \cite{polchinskipeet}
\begin{equation}
 \Lambda = \frac{1}{\sqrt{Ng^{2}_{YM}}}\left(\frac{r_{\max}}{\alpha^{\prime}}\right)^{\frac{5-p}{2}}, \label{eq:cutoff}
\end{equation}
$r_{\max}$ being the cut-off of $r$.  The effective coupling at an energy scale $E$ is defined by
\begin{equation}
 g^2_{eff}(E) = N g^{2}_{YM} E^{p-3}. \label{eq:geff}
\end{equation}

The bulk space has a curvature of \cite{maldacena}
\begin{equation}
 \alpha^{\prime}R \approx \sqrt{\frac{(r/\alpha^{\prime})^{3-p}}{Ng^2_{YM}} }. \label{eq:cur}
\end{equation}
Hence, for $p \neq 3$ both the dilaton and curvature of the bulk are not constant and depend on the radial coordinate of the bulk.
The approximate supergravity solution (\ref{eq:supgmetric}) remains valid as long as the curvature and the dilaton stay small
 \cite{maldacena}.  Therefore, when $p\neq 3$ the geometry (\ref{eq:supgmetric}) will only give a reliable holographic description
in a certain range of the radial direction. 
Later this will also put constraints on the sizes of regions for which we can reliably calculate the holographic entanglement entropy. 

  The constraint on the curvature is \cite{maldacena}
\begin{equation}
 \alpha^{\prime}R \ll 1,
\end{equation}
so that
\begin{equation}
 r \ll \alpha^{\prime}(Ng^2_{YM})^{\frac{1}{3-p}}. \label{eq:curlmt}
\end{equation}
The constraint on the dilaton is that \cite{maldacena}
\begin{equation}
 g_s\cdot e^{\phi} \ll 1,
\end{equation}
which leads to
\begin{equation}
 \left(\frac{r}{\alpha^{\prime}}\right)^{p-3} \ll g^{-2}_{YM}N^{\frac{p-3}{7-p}}. \label{eq:dillmt}
\end{equation}

\section{Calculation of entanglement entropy}

In this section we calculate the entanglement entropy of a region $A$ embedded in the boundary of the spacetime 
(\ref{eq:supgmetric}) on a fixed timeslice.  The edge of $A$, $\partial A$, will either be a sphere or a strip.  We will 
calculate the entanglement entropy of $A$ with equation (\ref{eq:pent}), by using an extremized surface $\gamma_A$ 
that connects with $\partial A$ at the boundary.
In these calculations we assume that $1 \leq p \leq 4$, since for $p \geq 5$ it is known that the duality is invalid 
\cite{polchinskipeet}, \cite{d5one}, \cite{d5two}.

\subsection{A strip on a $Dp$-brane boundary}

From formula (\ref{eq:pent}), the entanglement entropy of the strip is
\begin{equation}
S_{A} = \min_{\partial\gamma_A=\partial A}\frac{1}{4G^{(10)}_N}\int d\Omega_{8-p} d^{p-1}x \cdot
r^{7-p}_p\int^{\frac{\ell}{2}}_{-\frac{\ell}{2}}d x_1\cdot r\sqrt{\left(\frac{r}{r_p}\right)^{7-p} + \dot{r}^2}, \label{eq:pstrip}
\end{equation}
where $r$ now depends on $x_1$ in the induced metric on $\gamma_A$.

We can calculate the relation between $x_1$ and $r$ 
on the extremal surface as well as the minimal value of $r$ by analogy to the Hamiltonian formalism.
 It should be noted that in the coordinates we are using, the boundary space is located at $r \rightarrow \infty$, which is why 
the minimum $r$ corresponds to the apex of $\gamma_A$ in the bulk.
  Treating the integrand $I$ in (\ref{eq:pstrip}) the same as a Lagrangian in an action, and ignoring the prefactors of the integral,
 we find the analog of the Hamiltonian to be
\begin{equation}
h = \dot{r}\frac{\partial I}{\partial \dot{r}} - I
= -\frac{r^{8-p}}{r^{7-p}_p}\cdot\frac{1}{\sqrt{\left(\frac{r}{r_p}\right)^{7-p} + \dot{r}^2}}
\equiv -\frac{r^{8-p}_{\ast}}{r^{7-p}_p}\cdot\left(\frac{r_p}{r_{\ast}}\right)^{\frac{7-p}{2}},
\end{equation}
where $r_{\ast}$ is the minimum value of $r$ on the extremized $\gamma_A$.  The Hamiltonian analog $h$ only depends 
implicitly on $x_1$ and is therefore constant. We used the assumption that $\gamma_A$ is regular at $x_1=0$ and therefore
 $\dot{r}(x_1=0) = 0$ by symmetry.  Solving for $\dot{r}^2$, we have
\begin{equation}
 \dot{r}^2 = 
r^{-(7-p)}_pr^{7-p}_{\ast} \left(\frac{r}{r_{\ast}}\right)^{7-p}\left[\left(\frac{r}{r_{\ast}}\right)^{9-p}-1\right]. \label{eq:pz2}
\end{equation}
We can solve the differential equation to get the dependence of $r_{\ast}$ on $\ell$
 by integrating
\begin{equation}
 \int^{\frac{\ell}{2}}_{0} dx_1 = r^{\frac{7-p}{2}}_p r^{-\frac{5-p}{2}}_{\ast}\int^{\infty}_1d\left(\frac{r}{r_{\ast}}\right)\cdot 
\left(\frac{r}{r_{\ast}}\right)^{-\frac{7-p}{2}}\left[\left(\frac{r}{r_{\ast}}\right)^{9-p}-1\right]^{-\frac{1}{2}}. \label{eq:pzstar}
\end{equation}
Performing a change of variable we define 
$t = \frac{r_{\ast}}{r}$, so that equation (\ref{eq:pzstar}) becomes
\begin{eqnarray}
 \frac{\ell}{2} &=& r^{\frac{7-p}{2}}_p r^{-\frac{5-p}{2}}_{\ast}\int^1_0dt\cdot t^{6-p}[1-t^{9-p}]^{-\frac{1}{2}} \nonumber \\
&=& r^{\frac{7-p}{2}}_p r^{-\frac{5-p}{2}}_{\ast} 
\frac{2\sqrt{\pi}\Gamma\left(\frac{7-p}{9-p}\right)}{(5-p)\Gamma\left(\frac{5-p}{2(9-p)}\right)}. \label{eq:psqrt}
\end{eqnarray}
Hence we can solve for $r_{\ast}$:
\begin{equation}
 r_{\ast} = \left[\frac{4r^{\frac{7-p}{2}}_p}{\ell}
\cdot\frac{\sqrt{\pi}\Gamma\left(\frac{7-p}{9-p}\right)}{(5-p)\Gamma\left(\frac{5-p}{2(9-p)}\right)}\right]^{\frac{2}{5-p}}.
\end{equation}

Using equations (\ref{eq:pz2}) and (\ref{eq:psqrt}), we can write the minimized equation (\ref{eq:pstrip}) as 
\begin{equation}
 S_A = \frac{2\Omega_{8-p}}{4G^{(10)}_N} L^{p-1}r^{7-p}_p
r^2_{\ast}\int^{1}_{r_{\ast}/r_{\max}}dt \cdot t^{-3}\left[1-t^{9-p}\right]^{-\frac{1}{2}}, \label{eq:thintegral}
\end{equation}
where $t=\frac{r_{\ast}}{r}$ and $L$ is the length regulator of the strip.
We wish to extract the finite and infinite parts of the integral.
 Near $t=0$ (i.e. near the boundary of the space) we can write the integral as 
\begin{eqnarray}
&& \int_{{r_{\ast}/r_{\max}}}dt \cdot t^{-3}(1+\frac{1}{2}t^{9-p}+\dots) \nonumber \\
&=& \frac{1}{2}\left(\frac{r_{\max}}{r_{\ast}}\right)^{2} + \frac{1}{2(7-p)}\left(\frac{r_{\max}}{r_{\ast}}\right)^{p-7}
 + \mathcal{O}(r_{\max}^{-2(9-p)}).
\end{eqnarray}
The integral (\ref{eq:thintegral}) has a divergence of order $\mathcal{O}(r_{\max}^{2})$.  The finite part is the regularized integral
\begin{eqnarray}
&&\int^{1}_{0}dt \cdot t^{-3}\left[1-t^{9-p}\right]^{-\frac{1}{2}} - \frac{1}{2}\left(\frac{r_{\max}}{r_{\ast}}\right)^{2} \nonumber \\
&=& -\frac{\sqrt{\pi}}{2}\cdot\frac{\Gamma(\frac{7-p}{9-p})}{\Gamma(\frac{5-p}{2(9-p)})},
\end{eqnarray}
so that the (dimensionless) entanglement entropy of the strip on a $Dp$-brane is
\begin{eqnarray}
 S_{A} &=& \frac{1}{4G^{(10)}_N}\Omega_{8-p}L^{p-1}r^{7-p}_p
\left(r_{\max}^{2} - r^2_{\ast}\sqrt{\pi}\frac{\Gamma(\frac{7-p}{9-p})}{\Gamma(\frac{5-p}{2(9-p)})}\right) \label{eq:peq3ref}\\
&=&\frac{4\pi}{(2\pi)^{7}\alpha^{\prime 4}g^2_s} \Omega_{8-p}L^{p-1}(Ng^{2}_{YM} \alpha^{\prime 5-p} d_p)
\left(r_{\max}^{2} - r^2_{\ast}\sqrt{\pi}\frac{\Gamma(\frac{7-p}{9-p})}{\Gamma(\frac{5-p}{2(9-p)})}\right), \label{eq:sstripbefore}
\end{eqnarray}
after substituting in for the value of $G^{(10)}_N$ from (\ref{eq:newton}) and for $r^{7-p}_p$ from (\ref{eq:zp}).

We would like to rewrite our geometrically determined entanglement entropy in terms of quantities in the boundary gauge theory,  
i.e. in terms of the rank of the gauge group $N$, the UV cut-off $\Lambda$ as well the effective couplings $g^2_{eff}(E)$ 
at the relevant energy scales.

The entanglement entropy of the strip on the $Dp$-brane boundary theory then becomes
\begin{equation}
 S_{A} = A\cdot N^2\left(\Lambda L\right)^{p-1}\left(g^2_{eff}(\Lambda)\right)^{\frac{p-3}{5-p}} - 
B\cdot N^2\left(g^2_{eff}\left(\ell^{-1}\right)\right)^{\frac{p-3}{5-p}}\left(\frac{L}{\ell}\right)^{p-1}, \label{eq:entstrip}
\end{equation}
where the constants $A$ and $B$ are
\begin{eqnarray}
A &=& \frac{1}{(7-p)\pi}, \\
B &=& \frac{2^{\frac{22-4p}{5-p}}\pi^{\frac{17-2p}{10-2p}}}{(7-p)(5-p)^{\frac{4}{5-p}}}
 \left(\Gamma\left(\frac{7-p}{2}\right)\right)^\frac{2}{5-p}
\left(\frac{\Gamma(\frac{7-p}{9-p})}{\Gamma(\frac{5-p}{2(9-p)})}\right)^{\frac{9-p}{5-p}}.
\end{eqnarray}

Recall that in the case $p=3$, the metric (\ref{eq:supgmetric}) becomes that of AdS$_5 \times S^5$.  In that case, $r_3$ becomes the AdS curvature 
scale $R$ and the $5$-dimensional Newton's constant is given by $G^{(5)}_N=G^{(10)}_N/\Omega_{5}R$.  Introducing the short distance 
regulator $a = r^2_3/r_{\max}$ and substitute in for $r_{\ast}$ in the case $p=3$ in (\ref{eq:peq3ref}), the entanglement entropy of the strip corresponds 
exactly to the result of Ryu and Takayanagi for three dimensions in \cite{stripdisk}, namely
\begin{equation}
 S_A = \frac{R^3}{4 G^{(5)}_N} \left(\left(\frac{L}{a}\right)^2 - 4 \pi^{3/2}
\left(\frac{\Gamma\left(\frac{2}{3}\right)}{\Gamma\left(\frac{1}{6}\right)}\right)^{3}\left(\frac{L}{\ell}\right)^2\right).
\end{equation}
The expected EE of AdS/CFT is therefore reproduced in the specific case when the boundary field theory is conformal.  Also note in 
equation (\ref{eq:entstrip}), for $p=3$ the effective coupling disappears and the entanglement entropy of a CFT is therefore independent 
of its effective coupling.

The calculation breaks down at $p=5$ and likely for higher dimensions.  This is apparent from the fact that the 
 finite term in (\ref{eq:entstrip}) diverges when $p=5$.  In any event, the  holographic duality was also found to fail for $p=5$
 in \cite{d5one}, \cite{d5two} and \cite{polchinskipeet}.

We would like to find the constraints on the allowed widths $\ell$ of the strip for which the holography remains valid.
First, using the definition of $r_p$ in equation (\ref{eq:zp}), we can solve for $\alpha^{\prime}$ in terms of field theory quantities, namely 
\begin{equation}
 \alpha^{\prime} \approx \left(\frac{r^{7-p}_p}{Ng^2_{YM}}\right)^{\frac{1}{5-p}}.
\end{equation}

After substituting in for $\alpha^{\prime}$ for different values of $p$ into (\ref{eq:curlmt}) and (\ref{eq:dillmt}), 
we find that the constraints on $r$ can be written as
\begin{eqnarray}
 r^{3/2}_1N^{-1/3}(Ng^2_{YM})^{\frac{1}{4}} &\ll& r \ll r^{3/2}_1(Ng^2_{YM})^{\frac{1}{4}} \quad \textrm{when}\quad p=1, \\
r^{5/3}_2N^{-4/5}(Ng^2_{YM})^{\frac{2}{3}} &\ll& r  \ll r^{5/3}_2(Ng^2_{YM})^{\frac{2}{3}} \quad \textrm{when}\quad p=2, 
\label{eq:z2l} \\
0 &<& r < \infty \quad \textrm{when} \quad p=3\\
\frac{r^3_4}{(Ng^2_{YM})^2} &\ll& r \ll \frac{r^3_4N^{4/3}}{(Ng^2_{YM})^2} \quad  \textrm{when}\quad p=4 \label{eq:z4l}.
\end{eqnarray}
Note that the lower limits on $\frac{r}{\alpha^{\prime}}$ for $p=1$ and $2$ came from the dilaton's constraints while the lower limit 
on $\frac{r}{\alpha^{\prime}}$ when $p=4$ came from the curvature constraint and vice versa.  The reason for this inversion of roles 
comes from the fact that when $p<3$, the Yang-Mills coupling is relevant (it has units of inverse length), while for $p>3$, the 
Yang-Mills coupling is irrelevant (it has units of length).
  This also means that the coupling is relevant when $p<3$ and is large at low 
energy scales, while it is irrelevant when $p>3$ and therefore is large at high energies.
  Finally note that when $p=3$, the coupling is 
marginal (dimensionless) and the supergravity dual theory holds in the whole bulk, since
 this reproduces the AdS/CFT correspondence.

So we get the limits on the width of the strip to be
\begin{eqnarray}
 \frac{1}{\sqrt{Ng^2_{YM}}} &\ll& \ell \ll \frac{N^{\frac{2}{3}}}{\sqrt{Ng^2_{YM}}} \quad \textrm{when}\quad p=1, \label{eq:con1} \\
\frac{1}{Ng^2_{YM}} &\ll&  \ell  \ll \frac{N^{\frac{6}{5}}}{Ng^2_{YM}} \quad \textrm{when}\quad p=2, \\
0 &<& \ell < \infty \quad \textrm{when} \quad p=3,\\
\frac{Ng^2_{YM}}{N^{\frac{2}{3}}} &\ll& \ell \ll Ng^2_{YM} \quad  \textrm{when}\quad p=4. \label{eq:con4}
\end{eqnarray}

\subsection{A disk on the $Dp$-brane boundary}
\begin{figure}
 \def\svgwidth{10cm}
\centering
\input{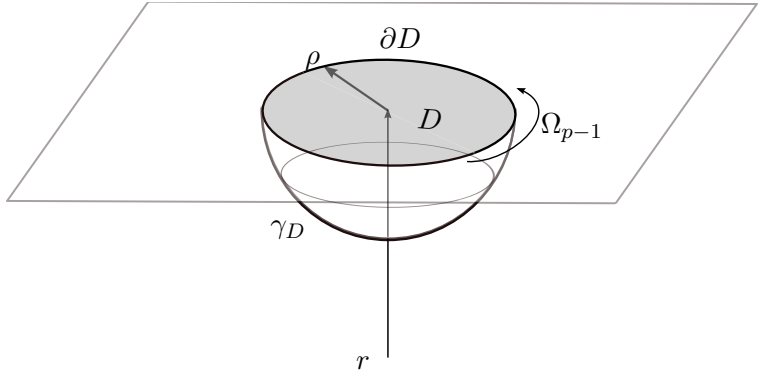}
\caption{The minimal surface $\gamma_D$ in the bulk AdS space connects with 
$\partial D$ on the $p$-dimensional boundary $Dp$-branes of the bulk supergravity solution.}\label{fig:dpbranedisk}
\end{figure}

We cannot solve exactly for the minimal surface to simplify the area integral.
  Instead, we  approximately solve a differential equation close to the boundary to solve for the minimal surface.
  In some cases we obtain a logarithmic divergence which will
 have a universal coefficient, that is a coefficient that does not depend on the regularization method.  We derive the general 
differential equation and solve it approximately for the cases $p=2$ and $p=4$.  
The case $p=3$ yields the same solution as one would get 
for $p=3$ in the AdS/CFT case as calculated in \cite{stripdisk},
 and $p=1$ yields the same result as for the strip in the previous subsection, since a disk and a strip 
on a line are the same.

The disk $D$ with a boundary $\partial D$ of radius $\ell$ (see figure \ref{fig:dpbranedisk})
is embedded in the boundary of the $10$-dimensional spacetime.  
Rewriting the spatial gauge theory on the boundary in (\ref{eq:supgmetric}) as
\begin{equation}
 \sum^p_{i=1}dx^2_i = d\rho^2 + \rho^2d\Omega^2_{p-1},
\end{equation}
we see that the induced metric on the bulk surface $\gamma_D$ is
\begin{equation}
 ds^2 = \left(\frac{r}{r_p}\right)^{\frac{7-p}{2}}\rho^2d\Omega^2_{p-1} +
 \left(\frac{r_p}{r}\right)^{\frac{7-p}{2}}\left[\left(\left(\frac{r}{r_p}\right)^{7-p}\dot{\rho}^2 + 1\right)dr^2 +
r^2d\Omega^2_{8-p}\right].
\end{equation}  
We wrote the metric in this form
 because we expect the minimal surface to be rotationally symmetric about the $r$-axis, since $\partial D$ is a sphere.
  Using this form of the metric and the string-frame equation for the entanglement entropy, we have
\begin{eqnarray}
 S_{D} &=& \min_{\partial\gamma_D=\partial D}\frac{4\pi}{(2\pi)^7\alpha^{\prime 4}g^2_s}
 \int_{\gamma_D}d^8x\sqrt{-g}e^{-2\phi} \nonumber \\
&=& \min_{\partial\gamma_D=\partial D} \frac{4\pi}{(2\pi)^7\alpha^{\prime 4}g^2_s}\Omega_{p-1}\Omega_{8-p}r^{11-p}_p
\int dz \frac{\rho^{p-1}}{z^3} \sqrt{\left(\frac{r_p}{z}\right)^{3-p}\dot{\rho}^2 + 1}, \label{eq:pdisk}
\end{eqnarray}
where  we made the change of variable $z = \frac{r^2_p}{r}$. 
The dot now indicates the derivative with respect to $z$.  In this new bulk radial coordinate the boundary of spacetime is located at $z=0$.

Using the usual method of treating the integral in (\ref{eq:pdisk}) like an action and the integrand as a 
Lagrangian, we can find a second order nonlinear differential equation from the Euler-Lagrange equation:
\begin{equation}
 2r^{2p}_p(p-1)\left(\frac{1}{z}\right)^p  + (9-p)r^6_p\left(\frac{1}{z}\right)^{7-p}\rho\dot{\rho}^3
- 2 \frac{r^{p+3}_p}{z^4}\left(-(p-1)z\dot{\rho}^2 + \rho[-(6-p)\dot{\rho} + z\ddot{\rho}]\right) = 0. \label{eq:difeq}
\end{equation}
We now solve the equation for specific values of $p$:

The case $\mathbf{p=1}$ is the same as the strip, as mentioned.

In the case $\mathbf{p=2}$ the differential equation (\ref{eq:difeq}) becomes
\begin{equation}
 2z^3 + 7r^2_2 \rho \dot{\rho}^3 - 2z r_2(-z\dot{\rho}^2 - 4\rho\dot{\rho} + z\rho\ddot{\rho}) = 0 \label{eq:difeq2}
\end{equation}
We will solve this differential equation close to the boundary ($z=0$) with a series representation for $r$ in terms of $z$:
\begin{equation}
 \rho(z) = \sum^{\infty}_{n=0} a_n z^n. \label{eq:series2}
\end{equation}

Because the differential equation is second order 
in derivatives of $z$, we need two  coefficients determined from the 
boundary conditions of the problem to fix the rest of the coefficients in the series.  
The coefficient $a_0 = \ell$ is one of these coefficients.  To determine the second such coefficient, we write
\begin{equation}
 \rho(z) = f(z) + a_{\alpha}z^{\alpha} + \mathcal{O}(z^{\alpha + 1}), \label{eq:r}
\end{equation}
where $a_{\alpha}$ is the lowest order coefficient that cannot be determined from $a_0$.  This means $a_{\alpha}$ needs to be determined 
from a boundary condition of the problem.  The function $f(z)$ is given by
\begin{equation}
 f(z) = \sum^{\alpha - 1}_{n = 0} a_n z^n.
\end{equation}
From the form of (\ref{eq:difeq2}), we can see that we do not need higher orders of $z$ to determine $a_{\alpha}$.

Inserting equation (\ref{eq:r}) into (\ref{eq:difeq2}),
we pick out the terms in the above expression that are lowest in order of $a_{\alpha}$ and $z$
 and only first order of $f$ and set their sum to zero, since we expect (\ref{eq:difeq2}) to equal zero at each order of the parameters.
  We therefore have 
\begin{equation}
z_pf(z)z^{\alpha}\left(8\alpha - 2\alpha(\alpha - 1)\right) = 0
\end{equation}
with solutions $\alpha = 0$ or $5$.
 We only need the terms up to order $z^3$ in the series (\ref{eq:series2}) to determine the divergent part of the entanglement entropy
and therefore do not need to know what coefficient $a_5$ ($a_{\alpha}$) is and only need one boundary condition.
  Substituting the series (\ref{eq:series2}) into 
(\ref{eq:difeq2}), we can solve for the first few coefficients:
\begin{eqnarray}
 a_0 = \ell,\quad a_3 = -\frac{1}{6r_2\ell}, \nonumber \\
a_1 = a_2 = a_4 = 0.
\end{eqnarray}

When $p=2$, entropy equation (\ref{eq:pdisk}) becomes
\begin{equation}
 S_{D} = \frac{4\pi}{(2\pi)^7\alpha^{\prime 4}g^2_s}\Omega_1\Omega_6 r^9_2
\int^{z_{\ast}}_{z_{\min}} dz \frac{\rho}{z^3} \sqrt{\left(\frac{r_2}{z}\right)\dot{\rho}^2 + 1}.
\end{equation}
Substituting in for $r$ and Taylor expanding the square-root, the integral above (ignoring the constant prefactors)  becomes
\begin{eqnarray}
 &&\int_{z_{\min}} dz\frac{\ell - \frac{z^3}{6r_2\ell} + \mathcal{O}(z^5)}{z^3}
\left(1 + \frac{1}{2}\cdot\frac{z^3}{9r_2\ell^2} + \mathcal{O}(z^9) \right) \nonumber \\
&=& \frac{1}{2}\ell(z_{\min})^{-2} + \mathcal{O}(z_{\min}) + \textrm{finite terms}.
\end{eqnarray}

Substituting in $z_{min} = r^2_2/r_{\max}$, $r^5_2 = Ng^2_{YM} \alpha^{\prime 3} d_2$, 
$\Omega_1 = 2\pi$, $\Omega_6 = \frac{16\pi^3}{15}$ and $d_2 = 6\pi^2$, as well as for the UV cut-off $\Lambda$ from (\ref{eq:cutoff})
 and the effective coupling $g^2_{eff}(\Lambda)$ from (\ref{eq:geff}),
the entanglement entropy can be written succinctly as
\begin{equation}
 S_{D} = \frac{N^2}{10\pi}(2\pi\ell)\Lambda\left(g^2_{eff}(\Lambda)\right)^{-\frac{1}{3}} + \textrm{finite terms}.
\end{equation}

Notice that the divergent term has a coefficient proportional to the area of $\partial D$ and the entropy therefore satisfies the 
area law. Also notice that the effective coupling is raised to the power
\begin{equation}
 \frac{p-3}{5-p} = -\frac{1}{3},
\end{equation}
the same power one gets for a strip on a $p$-brane.

In the case $\mathbf{p=3}$ the differential equation for the surface $\gamma_D$ becomes 
\begin{equation}
 2z(1 + \dot{\rho}^2) + \rho(3\dot{\rho} + 3\dot{\rho}^3 - z\ddot{\rho}) = 0,
\end{equation}
with the solution
\begin{equation}
 \rho^2 + z^2 = \ell^2.
\end{equation}
This is exactly the result we would expect for a disk on the boundary of AdS space.
  Since we have an exact result for $\rho(z)$, we can solve for
the entanglement entropy analytically.  By setting $r_3 = R$, the curvature scale of AdS, and using the definition of 
$G^{(5)}_N$ given in section $3.1$, we find that the entanglement entropy of the disk on a $D3$-brane is 
\begin{equation}
 S_D = -\frac{\pi R^3}{G^{(5)}_N}\log\left(\frac{\ell}{z_{\min}}\right) + \textrm{finite terms},
\end{equation}
which is the result we would get in a $(3+1)$-dimensional CFT \cite{stripdisk}.

In the case $\mathbf{p=4}$ the differential equation (\ref{eq:difeq}) becomes
\begin{equation}
 6r_4(r_4 + z\dot{\rho}^2) + \rho(4r_4\dot{\rho} + 5z\dot{\rho}^3 - 2r_4 z\ddot{\rho}) = 0. \label{eq:difeq4}
\end{equation}

Once again we will expand $r(z)$ as a series in $z$, close to the boundary $z=0$.  Just like in the $p=2$ case, from the fact
 that $\gamma_D$ meets $\partial D$ at the boundary space at $z=0$, we have that $a_0 = \ell$.
We again write
\begin{equation}
 \rho(z) = g(z) + a_{\alpha}z^{\alpha} + \textrm{higher order terms}.
\end{equation}
Substituting this into (\ref{eq:difeq4}), we set the expression to lowest order of $a_{\alpha}$, $z$ and $g$ to zero to find
\begin{equation}
 g(z)z^{\alpha-1}\left(4\alpha - 2\alpha(\alpha-1)\right) = 0,
\end{equation}
with solutions $\alpha=0$ or $3$.
  It turns out that the general solution of the series is
\begin{equation}
 \rho(z) = \sum^{\infty}_{m=0} a_m z^m + z^3\log(z)\sum^{\infty}_{n=0} b_n z^n,
\end{equation}
where the coefficients $a_1$, $a_2$ and $b_0$ are determined completely in terms of $a_0$,
 but the rest of the coefficients are determined in terms of both $a_0$ and $a_3$.
We only need to know the coefficients up to order $z^2$ to determine the divergent part of the entanglement entropy and therefore 
do not need an additional boundary condition to determine the coefficients
\begin{equation}
a_0=\ell,\quad a_1 = -\frac{3r_4}{2\ell},\quad  a_2 = -\frac{45r^2_4}{32\ell^3}.
\end{equation}

When the boundary space is four-dimensional the entropy formula (\ref{eq:pdisk}) becomes
\begin{equation}
 S_{D} = \frac{4\pi}{(2\pi)^7\alpha^{\prime 4}g^2_s}\Omega_3\Omega_4 r^7_4
\int dz \frac{\rho^{3}}{z^3} \sqrt{\left(\frac{z}{r_4}\right)\dot{r}^2 + 1}. \label{eq:4disk}
\end{equation}

Substituting in for $r$ and Taylor expanding around $z=0$ as before, the integral in (\ref{eq:4disk}) becomes
\begin{eqnarray}
 S_{D} = \frac{4\pi}{(2\pi)^7\alpha^{\prime 4}g^2_s}(2\pi^2)(\frac{8}{3}\pi^2)
\left[r^3_4\frac{\ell^3}{\delta^2} - r^6_4\frac{27\ell}{8\delta} 
+ r^9_4\frac{135}{128}\cdot\frac{1}{\ell}\log\left(\frac{\ell}{\delta}\right)\right],
\end{eqnarray}
after substituting in for $\Omega_3$ and $\Omega_4$.

Substituting in for the constants as well as the UV cut-off and the effective coupling as we did in the case $p=2$,
the entanglement entropy can be written in its final form of
\begin{equation}
 S_{D} = N^2g^2_{eff}(\Lambda)\left(\frac{1}{3\pi}(2\pi^2\ell^3)\Lambda^3 - \frac{9}{16}\ell \Lambda\right)
 + \frac{45}{2^{10}\pi}N^2g^2_{eff}\left(\ell^{-1}\right)\cdot\log(\ell \Lambda) + \textrm{finite terms}.
\end{equation}
The coefficient in front of the logarithm will be universal.
 Notice that the effective coupling is raised to the power
\begin{equation}
 \frac{p-3}{5-p} = 1,
\end{equation}
just like the case $p=2$ and the strip.

The calculated entanglement entropy will be valid only for a certain range of the radius $\ell$ of $D$, since the value 
of the maximum of $z$, namely $z_{\ast}$ will depend on $\ell$. While the finite part can become invalid for a too large $\ell$,
 the infinite 
part that was calculated here can only be invalidated if $\ell$ becomes too small.  This is because the high energy cutoff pertains 
to the part of the surface $\gamma_D$ close to the boundary space.  For $\ell$ too small, the entire surface will fall outside the 
allowed range for $z$, and even the high-energy cutoff will be too close to the boundary for holography to hold. 

Because the limits on $z$ in (\ref{eq:z2l}) and (\ref{eq:z4l}) ignore possible numerical coefficients, we expect that the same 
bounds will be valid for the disk.  The fact that the strip and disk are identical when $p=1$ hints that the $z$-$r$ profiles will not be 
radically different in the two cases when $p \neq 1$.  We can therefore reasonably expect the bounds on $\ell$ to be the same as
 equations (\ref{eq:con1}) to (\ref{eq:con4}), in particular the lower bounds which relate to the divergent part.

\section{Discussion}

The new results in this paper was that we expressed the entanglement entropy of a strip and disk in terms of the effective coupling 
at the relevant energy scales of the problem.

  There are three important
 similarities in the three calculated examples of the strip and disk on a $p$-brane (and the black brane in the appendix):
\begin{enumerate}
 \item There is a factor of $N^2$, which is the rank of the gauge group in the gauge theory dual
 to the supergravity solution.
\item The factor in the leading order divergence is proportional to the area of the boundary of the region being considered 
(the so-called area law).
\item The effective coupling of the gauge theory evaluated at the energy scales relevant to the problem
 is present and raised to the power $\frac{p-3}{5-p}$.
\end{enumerate}
The fact that these three characteristics show up in three distinct cases at the relevant energy scales suggest that these are 
general properties of holographically calculated entropy.

We further confirmed that the case of the gauge theory on a $D3$-brane is dual to the supergravity solution 
AdS$_5\times S^5$, and that
 the entanglement entropies of both the strip and disk on the $D3$-brane will correspond to the AdS/CFT 
calculations in \cite{stripdisk}.  It is important to note that the effective couplings present in 
the entanglement entropy solutions as well as thermal entropy in the case of the black brane in the appendix, are not present 
in the $D3$-brane case  due to the exponent of the effective coupling.  In the AdS/CFT calculations there weren't effective 
couplings present either.

An interesting difference between the AdS/CFT and $Dp$-brane entanglement entropies arose in the case of the disk.  In the CFT, we only 
get logarithmic divergences when $p$ is odd, but in the $Dp$-brane case we found a log-divergence when $p$=4.  
Therefore the gauge theory on a $Dp$-brane can give universal terms when $p$ is even.  This is similar to a recent result 
in \cite{hms}, in which the boundary CFT has a relevant deformation.

Important differences arose when we went from the $D3$-case to the more general case.  In contrast to the AdS/CFT duality 
in which calculations are believed to hold up to arbitrary length scales, the supergravity solution in the bulk 
obtains both a dilaton as well as a scalar curvature which depend on the distance one extends into the bulk.  
If either the curvature or the dilaton becomes too big, the supergravity solution becomes an inappropriate 
approximate solution \cite{maldacena}.
As we saw in section $3.1$ for the calculation of the strip on a $p$-brane, this places bounds on the size of the region for 
which we can reliably calculate the entanglement entropy.

\section*{Acknowledgements}
I would like to offer my sincere gratitude to R.C. Myers for his supervision and discussions throughout my project and clarifications 
on many points, without which I could not have completed it.
I would also like to thank N. Doroud,  A. Singh, A. Yale and especially
 A. Buchel, L-Y. Hung and M. Smolkin for valuable discussions during this project.
  I thank C. Durnford for helpful discussions and for helping me with the drawing of the figures.
I acknowledge the Perimeter Scholars International program
 for making this opportunity possible with funding and for providing me with an 
academically enriching year.  Research at Perimeter Institute is supported by the Government of Canada
through Industry Canada and by the Province of Ontario through the Ministry of Research
and Innovation.

\appendix
\section{Thermal entropy of a black brane}

We would like to compare the entanglement entropy calculated in section $3$ with the thermal entropy of the same gauge theory, to 
see if it shows the same behaviour in terms of field theory quantities.  To raise the gauge theory to a finite temperature, 
 we add a black hole to the bulk space.  The black hole's temperature will be the relevant energy scale of the gauge theory and 
 we will use the effective coupling at this energy scale, as per equation (\ref{eq:geff}).

The black hole metric in the supergravity solution (modified from \cite{maldacena} to fit our conventions) is 
\begin{equation}
ds^2 = \left(\frac{r}{r_p}\right)^{\frac{7-p}{2}}\left[-\left(1-\left(\frac{r_0}{r}\right)^{7-p}\right)dt+dx^2_{\parallel}\right]
 + \left(\frac{r_p}{r}\right)^{\frac{7-p}{2}}\frac{dr^2}{\left(1-\left(\frac{r_0}{r}\right)^{7-p}\right)} +
r^{\frac{7-p}{2}}_p \cdot r^{\frac{p-3}{2}}d\Omega^2_{8-p}, \label{eq:bhmetric}
\end{equation}
where $r_0$ is the position of the black brane in the bulk and is given by
\begin{equation}
 r^{7-p}_0 = a_p g^4_{YM} \varepsilon \alpha^{\prime 7-p}; \quad
a_p = \frac{2^{11-2p}\pi^{\frac{13-3p}{2}}}{9-p}\Gamma\left(\frac{9-p}{2}\right)
\end{equation}
and $\varepsilon$ is the energy density in the field theory (with units of energy divided by length to the power $p$).  
Since the entropy of the black hole
\begin{equation}
 S = \frac{A_{hor}}{4G^{10}_N}
\end{equation}
is divergent (the area is infinite) we will work with its entropy density as given in \cite{maldacena},
\begin{eqnarray}
 s&\equiv&\frac{S}{\textrm{Vol}_{BH}} \\
&=& \frac{2^{\frac{3}{2}}a^{\frac{1}{7-p}}_p}{(7-p)^{\frac{1}{2}}(9-p)^{\frac{1}{2}}}
g^{2\cdot\frac{p-3}{2(7-p)}}_{YM}\sqrt{N}\varepsilon^{\frac{9-p}{2(7-p)}}. \label{eq:endens}
\end{eqnarray}

In order to find the temperature of the black hole we consider the singular parts of the metric, namely the $t$ and $r$ components 
in the near-horizon limit of the Euclideanized metric (i.e. $t\rightarrow t_E = it$).  
That means that we want to write the $t-r$ part of the metric in a form
\begin{equation}
 ds^2_{\textrm{near hor}} = \rho^2 d\tau^2 + d\rho^2. \label{eq:polar}
\end{equation}

Then the Euclidean radial coordinate $\rho$ turns out to be
\begin{equation}
 \rho = 2\sqrt{\frac{r-r_0}{7-p}}r^{\frac{7-p}{4}}_pr^{\frac{5-p}{4}}_0
\end{equation}
and the Euclidean time $\tau$ is
\begin{equation}
 \tau = 2\pi T_{BH}t_E,
\end{equation}
where $T_{BH}$ is the temperature of the black hole
\begin{equation}
 T = \frac{(7-p)(a_pg^4_{YM}\varepsilon)^\frac{5-p}{2(7-p)}}{4\pi\sqrt{Ng^2_{YM}d_p}}.
\end{equation}

Solving for $\varepsilon$ and substituting it back in the expression for the entropy density, the expression simplifies to
\begin{equation}
 s = \frac{2^{\frac{3}{2}}a^{-\frac{1}{2}}_pd_p^{\frac{9-p}{2(5-p)}}}{(7-p)^{\frac{1}{2}}(9-p)^{\frac{1}{2}}}
\left(\frac{4\pi}{7-p}\right)^{\frac{9-p}{5-p}}\left(N^{7-p}(g^2_{YM})^{p-3}T^{9-p}\right)^\frac{1}{5-p}.
\end{equation}

We can further simplify the expression by writing it in terms of the effective coupling at the energy scale
 of the black hole (the temperature), namely
\begin{equation}
 g^2_{eff}(T) = Ng^2_{YM} T^{p-3}.
\end{equation}
The black hole entropy density is then
\begin{equation}
s = \frac{2^{\frac{3}{2}}a^{-\frac{1}{2}}_pd_p^{\frac{9-p}{2(5-p)}}}{(7-p)^{\frac{1}{2}}(9-p)^{\frac{1}{2}}}\left(\frac{4\pi}{7-p}\right)^{\frac{9-p}{5-p}}
N^2 T^p (g^2_{eff}(T))^{\frac{p-3}{5-p}}. 
\end{equation}


\begin{thebibliography}{99}

\bibitem{maldacena1} J. M. Maldacena, ``The large N limit of superconformal field theories and supergravity,''
Adv. Theor. Math. Phys. \textbf{2}, 231 (1998) [Int. J. Theor. Phys. \textbf{38}, 1113 (1999)] [arXiv:hep-th/9711200].

\bibitem{gubser} S. S. Gubser, I. R. Klebanov and A. M. Polyakov, ``Gauge theory correlators from noncritical
string theory,'' Phys. Lett. B \textbf{428}, 105 (1998) [arXiv:hep-th/9802109].

\bibitem{witten} E. Witten, ``Anti-de Sitter space and holography,'' Adv. Theor. Math. Phys. \textbf{2}, 253 (1998)
[arXiv:hep-th/9802150].

\bibitem{aharony} O. Aharony, S. S. Gubser, J. M. Maldacena, H. Ooguri and Y. Oz, ``Large N field theories,
string theory and gravity,'' Phys. Rept. \textbf{323}, 183 (2000) [arXiv:hep-th/9905111].

\bibitem{mcgreevy} J. McGreevy, ``Holographic duality
with a view toward many-body physics,'' arXiv:0909.0518 [hep-th] (2010).

\bibitem{polchinski} J. Polchinski, ``Introduction to gauge/gravity duality,'' arXiv:1010.6134 [hep-th] (2010).

\bibitem{malchapter} J. M. Maldacena, ``The gauge/gravity duality,'' arXiv:1106.6073 [hep-th] (2011).

\bibitem{maldacena} N. Itzhaki, J. M. Maldacena, J. Sonnenschein and S. Yankielowicz,
``Supergravity and the large N limit of theories
with sixteen supercharges,'' Phys. Rev. D \textbf{58}, 046004 (1998) [arXiv:hep-th/9802042].


\bibitem{bh} J. D. Bekenstein, ``Black holes and entropy,'' 
Phys. Rev. D \textbf{7}, 2333 (1973).

\bibitem{hawk} S. W. Hawking, ``Particle creation by black holes,''
 Commun. Math. Phys. \textbf{43}, 199 (1975) [Erratum-ibid. \textbf{46}, 206 (1976)].

\bibitem{rt} S. Ryu and T. Takayanagi, ``Holographic derivation of entanglement entropy from AdS/CFT,'' 
Phys. Rev. Lett. \textbf{96}, 181602 (2006) [arXiv:hep-th/0603001].

\bibitem{stripdisk} S. Ryu and T. Takayanagi, ``Aspects of holographic entanglement entropy,'' 
JHEP \textbf{0608}, 045 (2006) [arXiv:hep-th/0605073].

\bibitem{klebanov} I. R. Klebanov, D. Kutasov and A. Murugan, ``Entanglement as a probe of confinement,'' 
Nucl. Phys. B \textbf{796}, 274-293 (2008) [arXiv:0709.2140 [hep-th]].

\bibitem{pakman} A. Pakman and A. Parnachev, ``Topological entanglement entropy and holography,'' 
JHEP \textbf{0807}, 097 (2008) [arXiv:0805.1891 [hep-th]]

\bibitem{d3} D. Arean, P. Merlatti, C. Nunez and A. V. Ramallo, ``String duals of two-dimensional (4,4) supersymmetric gauge theories,''
 JHEP \textbf{0812}, 054 (2008) [arXiv:0810.1053 [hep-th]]

\bibitem{d4} A. V. Ramallo, J. P. Shock and D. Zoakos, ``Holographic flavor in N=4 gauge theories in 3d from wrapped branes,'' 
 JHEP \textbf{0902}, 001 (2009) [arXiv:0812.1975 [hep-th]]


\bibitem{Asplund:2011cq}
  C. T. Asplund and S. G. Avery,
  ``Evolution of entanglement entropy in the D1-D5 brane system,''
    [arXiv:1108.2510 [hep-th]].

\bibitem{conventions} R. C. Myers and R. M. Thomson, ``Holographic mesons in various dimensions,'' 
JHEP \textbf{0609}, 066 (2006) [arXiv:hep-th/0605017].

\bibitem{polchinskipeet} A.W. Peet and J. Polchinski, ``UV/IR relations in AdS dynamics,'' Phys. Rev. D \textbf{59}, 065011 (1998)
 [arXiv:hep-th/9809022].

\bibitem{skenderis} I. Kanitscheider, K. Skenderis, M. Taylor, ``Precision holography for non-conformal branes,'' JHEP \textbf{09},
094 (2008) [arXiv:0807.3324 [hep-th]].

\bibitem{universal} R. C. Myers, A. Sinha, ``Holographic c-theorems in arbitrary dimensions,'' 
JHEP \textbf{1101}, 125 (2011) [arXiv:1011.5819 [hep-th]].


\bibitem{hms} L-Y. Hung, R. C. Myers and M. Smolkin, ``Some calculable contributions to holographic entanglement entropy,''
 arXiv:1105.6055 [hep-th] (2011).

\bibitem{d5one} O. Aharony, M. Berkooz, D. Kutasov, and N. Seiberg, ``Linear dilatons, NS five-branes and holography,'' 
JHEP \textbf{9810}, 004 (1998) [arXiv:hep-th/9808149].

\bibitem{d5two} J. L. F. Barbon and E. Rabinovici, ``Extensivity versus holography in
anti-de Sitter spaces,'' Nucl. Phys. B \textbf{545}, 371-384 (1999) [arXiv:hep-th/9805143].

\end{thebibliography}
\end{document}